\documentclass[sigconf,nonacm]{acmart}
\usepackage{enumitem}
\usepackage{tcolorbox}
\tcbuselibrary{breakable}
\AtBeginDocument{%
  }

\setcopyright{none}
\setlist[itemize]{noitemsep,leftmargin=*,topsep=0pt}

\begin{document}

\title[Do GEO Gains Survive the Live Web?]
{Do GEO Gains Survive the Live Web? Evaluating In-the-Wild Visibility Gains of Generative Engine Optimization via LIVE-GEO}

\author{Runze Kong}
\affiliation{%
  \department{Electrical Engineering and Computer Science}
  \institution{University of Michigan, Ann Arbor}
  \city{Ann Arbor}
  \country{USA}}

\author{Haibo Jin}
\affiliation{%
  \department{School of Information Sciences}
  \institution{University of Illinois Urbana-Champaign}
  \city{Champaign}
  \country{USA}}

\author{Xucheng Yu}
\affiliation{%
  \institution{University of Illinois Urbana-Champaign}
  \city{Champaign}
  \country{USA}}

\author{Xiaoqi Wang}
\affiliation[obeypunctuation=true]{%
  \institution{Starc Institute}
  \city{}
  \country{}}

\author{Huimin Zeng}
\affiliation{%
  \department{School of Information Sciences}
  \institution{University of Illinois Urbana-Champaign}
  \city{Champaign}
  \country{USA}}

\author{Haohan Wang}
\affiliation{%
  \institution{University of Illinois Urbana-Champaign}
  \city{Champaign}
  \country{USA}}
\email{haohanw@illinois.edu}

\renewcommand{\shortauthors}{Kong et al.}
\begin{abstract}
Generative engine optimization (GEO) rewrites web content so that LLM-based search systems, also termed generative engines (GEs), rank it higher in their answers and recommendations. However, existing GEO evaluations typically rely on directly supplied text or controlled retrieval pipelines, and whether GEO gains survive the live Web remains unclear, leaving a gap in understanding how GEO strategies affect the visibility of live webpages when accessed and ranked by GEs. To address this gap, we introduce LIVE-GEO, a benchmark for \textbf{L}ive, \textbf{I}n-the-wild \textbf{V}isibility \textbf{E}valuation for GEO. LIVE-GEO publishes GEO-rewritten products as live webpages. We use LIVE-GEO to evaluate five existing GEO strategies (CORE, C-SEO, E-GEO, AgenticGEO, and Statistics Addition), applying them to the product descriptions and reviews of 80 fictional brands across 18 product families. We run experiments on three \textbf{GEs} (ChatGPT with GPT-6, Gemini 3.6 Thinking, and Grok in Expert mode), four direct-text \textbf{API evaluators} (GPT-4o, DeepSeek, Kimi, and Claude), and a direct-text \textbf{local evaluator} (Qwen2.5-3B-Instruct). With frozen $\pm0.2\%$ prices, most direct-text evaluators and GEs receiving product URLs give CORE the highest Top-1 rate, reaching 66.3\%--76.3\% on the GEs, yet other strategies vary in relative performance, and the GEs select the same top product for only 48.8\%--66.3\% of brands. We also find that ChatGPT is sensitive to the order of candidate URLs: CORE's Top-1 rate is 73.8\% and 93.8\% under two opposite fixed orders, and the randomized order used by LIVE-GEO yields 82.5\%, between the two. In the numerical-faithfulness check, the JEV verifier agrees with the reference labels on 90.7\% of 194 numerical benchmark cases and with human judgments on all 50 sampled descriptions, and it flags 86 of the 400 rewritten descriptions, including all 80 Statistics Addition descriptions.

\end{abstract}


\begin{teaserfigure}
  \centering
 \includegraphics[width=1\textwidth]{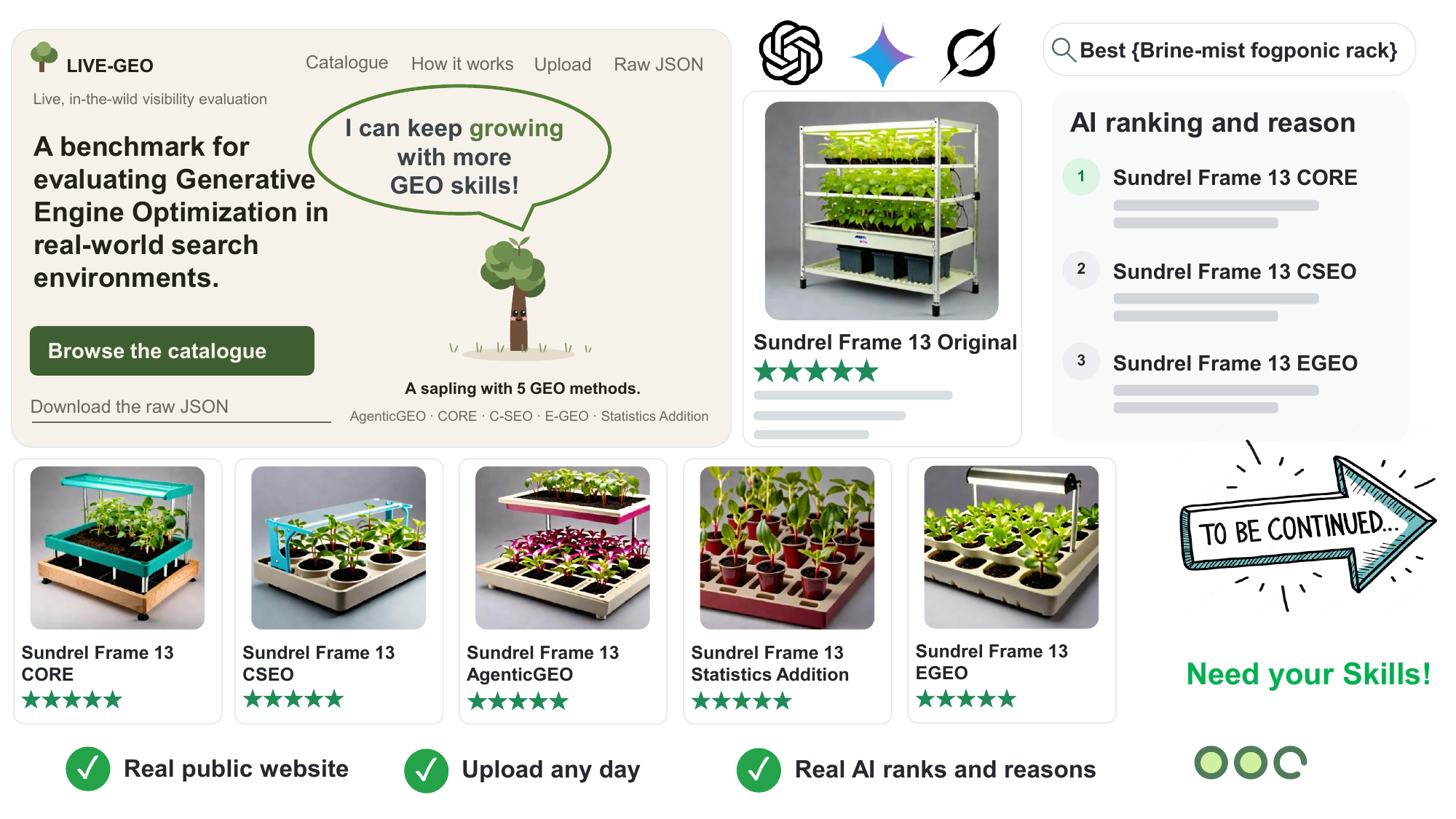}
  \caption{LIVE-GEO publishes Original and rewritten products as live webpages for evaluation through GEs. The interface and rankings are illustrative; the study covers 80 brands.}
  \Description{Illustrative LIVE-GEO interface showing a public website, Original and rewritten product listings, and a ranking returned by deployed AI products.}
  \label{fig:teaser}
\end{teaserfigure}


\maketitle

\section{Introduction}
\label{sec:introduction}
Generative engine optimization (GEO) aims to enhance the visibility of web content in responses generated by large language model (LLM)-based search systems, also called generative engines (GEs) \citep{aggarwal2024geo}. GEs combine web retrieval with an LLM that synthesizes responses from retrieved content \citep{aggarwal2024geo,lewis2020rag,nakano2021webgpt,gao2023ragsurvey}. 

Prior GEO strategies have mainly been evaluated either by supplying text directly to an LLM or through controlled retrieval pipelines built by researchers. For example, GEO-Bench, C-SEO Bench, and E-GEO place the rewritten content in the LLM input together with a fixed set of competing sources and measure its visibility or rank in the generated response \citep{aggarwal2024geo,puerto2025cseo,bagga2026egeo}. Because the candidate set is fixed in advance, these benchmarks evaluate selection but not retrieval. Other benchmarks also simulate the retrieval stage. For instance, SAGEO Arena and Counter-GEO-Bench build their own retrieval, reranking, and generation stages over collected web documents, so rewritten content is also evaluated at retrieval \citep{kim2026sageo,zheng2026countergeo}. Both settings are reproducible and easy to use, which supports controlled comparisons between GEO strategies. However, neither setting evaluates rewritten pages published on the live Web, even when it simulates the retrieval stage. In contrast, GEs retrieve and select web content through black-box pipelines \citep{aggarwal2024geo}, and different GEs draw on different sources \citep{kirsten2026characterizingwebsearch}. GEO gains measured in controlled settings may therefore not transfer to GEs, and whether these gains survive when GEs access the rewritten pages on the live Web remains unclear. 

To address this gap, we introduce LIVE-GEO (\textbf{L}ive, \textbf{I}n-the-wild \textbf{V}isibility \textbf{E}valuation for GEO), a benchmark that publishes GEO-rewritten products on the live Web and observes how GEs rank them. LIVE-GEO covers 80 fictional brands across 18 product families. It runs as the public product site of a startup partner, so its pages are reachable by GEs in the same way as other commercial sites. The products of a brand share the same specifications and differ only in their descriptions, reviews, and a small price offset: an Original product generated from templates and rewritten products, each produced by one GEO strategy, so differences in a GE's ranking of them come from the rewritten text rather than from the product itself. Users download the Original descriptions and reviews, rewrite them with their own GEO strategy, and upload the result. After our safety screening and numerical-faithfulness verification, the submission is published on our website as a new rewritten product, and users can then obtain its ranking from GEs.

We use LIVE-GEO to evaluate five existing GEO strategies, implemented from their released prompts or code. Specifically, we investigate four research questions (RQs) on LIVE-GEO: \textbf{RQ1:} How consistent is GEO visibility across GEs and direct-text evaluators? \textbf{RQ2:} How sensitive is visibility to repeated evaluation, candidate order, and public identity? \textbf{RQ3:} How do small price differences and URL entry conditions affect observed visibility? \textbf{RQ4:} How reliable are numerical-faithfulness flags, and how does applying them affect visibility? We answer all four in Section~\ref{sec:results}.

We compare three GEs used through their web interfaces, ChatGPT (GPT-6), Gemini (3.6 Thinking), and Grok (Expert mode), with five direct-text evaluators: four API models (GPT-4o, DeepSeek-V4-Pro, Kimi K2.6, and Claude Sonnet 5.5) and a local model (Qwen2.5-3B-Instruct). With frozen $\pm0.2\%$ prices and product URLs supplied to the GEs, CORE achieves the highest Top-1 rate with seven of the eight evaluators, including rates of 66.3\%--76.3\% on the GEs, yet the GEs agree on the top product for only 48.8\%--66.3\% of brands, and agreement between direct-text evaluators and GEs ranges from Spearman $\rho=0.50$ to $\rho=-0.31$. CORE's Top-1 rate on ChatGPT ranges from 73.8\% to 93.8\% across opposite fixed orders and differs between the unmasked repeat (83.8\%) and  masked run (76.3\%), whereas a repeat with the same presentation is stable ($\rho=0.85$), and price or URL-entry changes affect the values but not the Top-1 leader. The numerical-faithfulness verifier agrees with the reference labels on 90.7\% of 194 benchmark cases and with the human majority on all 50 reviewed descriptions.

\section{Related Work}
\label{sec:related-work}

\textbf{GEO Content Optimization Strategies.}\label{sec:geo-rewriting}
Heuristic GEO strategies rely on manually designed rules or criteria to guide content rewriting. Statistics Addition incorporates quantitative statistics into the content, while Cite Sources adds relevant source citations \citep{aggarwal2024geo}. The Content Improvement strategy of C-SEO Bench (C-SEO in this paper) combines multiple such transformations into a single rewriting strategy \citep{puerto2025cseo}. By contrast, feedback-based GEO strategies derive optimization signals from GE responses. AutoGEO learns GE preferences from how documents are used in generated responses and applies these preferences to prompt-based rewriting and reinforcement learning \citep{wu2026autogeo}. AgenticGEO further approximates such preference feedback with a surrogate critic and combines it with self-evolving strategy search \citep{yuan2026agenticgeo}. E-GEO iteratively optimizes rewriting prompts to improve product rankings \citep{bagga2026egeo}. Strategic text sequences (STS) and StealthRank study ranking-oriented adversarial optimization \citep{kumar2024manipulatingllms,tang2026stealthrank}. CORE's query-based optimization revises appended content according to differences between observed and target rankings \citep{jin2026core}.

\textbf{Evaluation Settings for GEO.}\label{sec:evaluation-settings}
GEO-Bench, C-SEO Bench, and E-GEO compare rewrites within preselected candidate sets, measuring visibility, citations, or rank once the candidates are fixed \citep{aggarwal2024geo,puerto2025cseo,bagga2026egeo}. As in fixed-data IR benchmarks such as MS MARCO and BEIR, shared queries and sources support controlled comparisons \citep{nguyen2016msmarco,thakur2021beir}. Counter-GEO-Bench retrieves and reranks within five preselected sources per query to assess defenses against distorting rewrites \citep{zheng2026countergeo}. SAGEO Arena reindexes rewrites in a crawled corpus and tracks visibility through retrieval, reranking, and generation, revealing that generation-stage gains can be lost upstream \citep{kim2026sageo}. Work on live GEs is of three kinds: audits ask whether GE answers cite verifiable sources \citep{liu-etal-2023-evaluating,venkit2025answerengines,venkit2026deeptrace}, user studies ask how people search with GEs \citep{miroyan2026searcharena,kirsten2026characterizingwebsearch}, and one-off experiments show that injected page content can change the rankings a GE returns \citep{pfrommer-etal-2024-ranking,nestaas2025adversarialseo}; none of them offers a standing benchmark on which GEO strategies can be compared. Outside GEO, live benchmarks keep evaluation current by refreshing test items against contamination, as in LiveBench and LiveCodeBench \citep{sainz-etal-2023-nlp,white2025livebench,jain2025livecodebench}, or by continually collecting human judgments, as in Chatbot Arena \citep{chiang2024chatbotarena}. LIVE-GEO brings this live design to GEO.

\textbf{Numerical Faithfulness in GEO.}\label{sec:faithfulness-constraint}
Prior work has addressed factual errors through factual-consistency detection, such as MiniCheck \citep{tang-etal-2024-minicheck}, and verification-based revision methods, including Chain-of-Verification and RARR \citep{dhuliawala-etal-2024-chain,gao-etal-2023-rarr}. Claim-verification datasets over tables and text, including TabFact, FEVEROUS, and SciTab \citep{Chen2020TabFact:, aly2021feverous,lu-etal-2023-scitab}, pair numerical claims with structured evidence; we draw the test cases for our verifier from them (Table~\ref{tab:jev-benchmark}).

\textbf{Key Differences.}
LIVE-GEO differs from prior work in three respects. Unlike fixed-source and retrieval-pipeline benchmarks, it publishes Original and rewritten products on the live Web, so GEs retrieve and rank the pages themselves. Unlike one-off live tests, it is a standing benchmark with frozen snapshots, randomized candidate order, and recorded identity and price settings, so measurements by different users are comparable. Unlike verification methods that revise text, it flags numerical claims against product specifications (Section~\ref{sec:faithfulness}) without changing the text or the observed rankings.

\section{LIVE-GEO Benchmark}
\label{sec:method}

\begin{figure*}[t]
  \centering
  \includegraphics[width=1 \textwidth,keepaspectratio]{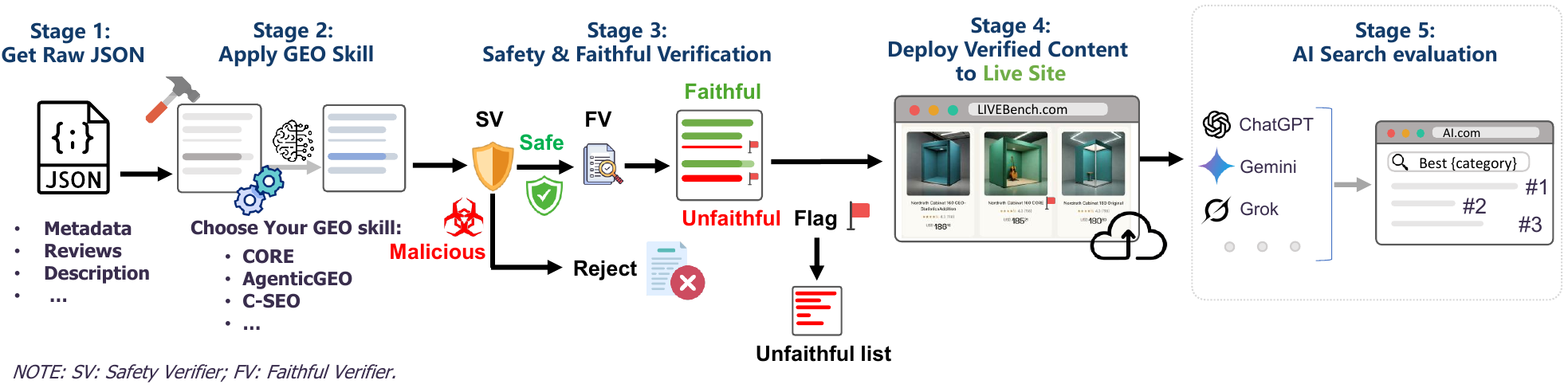}
  \vspace{-20pt}
  \caption{The five stages of LIVE-GEO: download product records, apply a GEO strategy, screen submissions, publish accepted products, and evaluate through GEs. Safety failures are rejected; numerical-faithfulness flags are recorded separately without deleting flagged text or changing observed rankings. The displayed ranks are illustrative.}
  \Description{Users download raw JSON and apply a GEO strategy. Safety screening rejects unsafe submissions; numerical verification flags potentially unsupported claims in descriptions. Accepted products are published, and users evaluate them through ChatGPT, Gemini, or Grok. Flags are exported separately.}
  \label{fig:live-workflow}
\end{figure*}

LIVE-GEO tests whether GEO gains survive on the live Web. It publishes rewritten products as live webpages and lets GEs such as ChatGPT and Gemini rank them. Each brand has an Original product, and each submitted GEO rewrite becomes a new rewritten product of the same brand with the same specifications. We measure a product's visibility by its position in the ranking that a GE returns for a category query such as best $X$, and a strategy's GEO gain by the visibility of its rewritten products relative to the other products of the same brand. Figure~\ref{fig:live-workflow} shows the five stages, which the subsections below describe in turn.

\subsection{Catalogue of Fictional Products}
\label{sec:catalogue}
We construct fictional products and brands in 80 uncommon categories to reduce interference with existing brands and webpages. These categories span 18 product families (Appendix Table~\ref{tab:families}), with one brand per category and six predefined specification attributes per family. A Python generator uses a fixed random seed to sample specification values and insert them into predefined templates, producing an Original description and seven reviews grounded in those specifications. All products within a brand share the same non-price specifications, enabling comparisons of products that differ only in their descriptions, reviews, and a small price offset.

Users download a public raw JSON file containing category and brand information, non-price specifications, a description, and seven reviews with their identifiers, titles, text, authors, dates, and ratings. Only the description and review titles and text may be rewritten. After accepting an upload, the platform assigns the rewritten product a price drawn at random within $\pm0.2\%$ of the brand's base price and names it by combining the brand name with the uploaded filename.

\subsection{Rewriting and Submission}
\label{sec:submission}
Users may apply any GEO strategy to the editable JSON fields and upload one file per strategy at any time, without an account, email, or invitation. After an initial submission is accepted, the platform issues a receipt token for updates under the same filename before a snapshot is frozen; new strategies require an unused filename. Submissions undergo two checks: a lightweight code gate rejects malformed or incomplete files, changes to read-only fields, and prohibited content such as HTML, scripts, links, images, or hidden characters before storage; a separate JEV safety screen identifies and rejects potentially malicious content targeting the platform, its visitors, or AI systems, and we then review each remaining file manually. Files that pass these checks proceed to numerical-faithfulness verification (Section~\ref{sec:faithfulness}). 

\subsection{Faithfulness Verification}
\label{sec:faithfulness}
We define \emph{numerical faithfulness} as requiring concrete numerical claims in product descriptions to be consistent with and supported by the product specifications. A description violates this policy if it contradicts a specified value or introduces unsupported numerical information. Equivalent unit conversions and values directly derivable from the specifications are considered faithful. Non-numerical claims are outside the scope of this check.

After safety screening and manual review, JEV \citep{deusser2026jev} checks each submitted description against its product specifications before publication. We choose JEV because it supports fixed-choice classification with associated probabilities, fitting our policy-based verification task without generating replacement text. The same verification policy is applied across GEO strategies. This check covers rewritten descriptions only; Original descriptions and reviews are excluded.

Descriptions identified as potentially violating the numerical policy receive advisory flags. These flags do not cause rejection, text deletion, or automatic ranking changes. We export flagged products to a separate JSON file, leaving product-page content unchanged. Users who apply a ranking penalty can place flagged products after unflagged ones while preserving the relative order within each group.

\subsection{Publication in Frozen Snapshots}
\label{sec:publication}
LIVE-GEO publishes accepted submissions in frozen snapshots. Before a snapshot is frozen, users may submit new GEO strategies or revise their uploaded text at any time; once it is frozen, its content no longer changes. A snapshot fixes the candidate products, content versions, public names, prices, and page-building configuration, so a product's rank does not shift because competitors were added or updated after the freeze. The first snapshot includes six products for each brand (one Original and five rewritten by the GEO strategies evaluated in this paper), each with a separate URL. A snapshot carries an identifier and its freeze time, which users record together with the capture time of each measurement, so that a ranking can be traced back to the exact products and text it was measured on.

\subsection{Live Observation}
\label{sec:live-observation}
Users evaluate published products through GEs, for example by asking for the best product in a category and requesting a ranking of the candidates. LIVE-GEO supports two query entry points: a brand-page URL, from which the GE is asked to discover and open individual products, or the product-page URLs supplied directly. To keep a fixed candidate order from favoring any product, the brand page lists its products in a new random order on every request; when supplying product URLs, we shuffle them once per brand and retain that order across repeated measurements. Users may design their own prompts. Our study uses a fixed template within each experimental condition and evaluates ChatGPT, Gemini, and Grok. Appendix~\ref{app:prompts} provides the prompts. 
\section{Experiments}
\label{sec:experiments}
\setcounter{topnumber}{4}
\setcounter{bottomnumber}{4}
\setcounter{totalnumber}{6}
\setcounter{dbltopnumber}{4}
\renewcommand{\topfraction}{0.95}
\renewcommand{\bottomfraction}{0.95}
\renewcommand{\dbltopfraction}{0.95}
\renewcommand{\textfraction}{0.05}
\renewcommand{\floatpagefraction}{0.85}
\renewcommand{\dblfloatpagefraction}{0.85}

We examine whether the same GEO rewrites receive similar rankings across evaluators and experimental conditions. We compare rankings produced by three GEs (ChatGPT, Gemini, and Grok) with those of five direct-text evaluators (Table~\ref{tab:ranks}). We address four research questions:

\textbf{RQ1:}\label{rq:consistency}
How consistent is GEO visibility across GEs and direct-text evaluators?

\textbf{RQ2:}\label{rq:sensitivity}
How sensitive is visibility to repeated evaluation, candidate order, and public identity?

\textbf{RQ3:}\label{rq:price-url}
How do small price differences and URL entry conditions affect observed visibility?

\textbf{RQ4:}\label{rq:faithfulness}
How reliable are numerical-faithfulness flags, and how does applying them affect visibility?

\subsection{Experimental Setup}
\label{sec:experimental-setup}

\textbf{Dataset.} We evaluate five GEO strategies on the 80 brands. Each brand has an Original product and five rewritten products produced by Statistics Addition \citep{aggarwal2024geo}, C-SEO \citep{puerto2025cseo}, E-GEO \citep{bagga2026egeo}, AgenticGEO \citep{yuan2026agenticgeo}, and CORE \citep{jin2026core}; each product shows a description and seven reviews, so every ranking compares six candidates.

\textbf{GEO baselines.} We generate the descriptions with the methods' released prompts or code and apply the same rewriting instructions to the review titles and bodies, keeping the specifications and the review identities, order, and ratings unchanged. The four non-CORE strategies use GPT-4.1. For CORE, we keep the authors' ranking-feedback and suffix-optimization loop: GPT-3.5 Turbo writes the initial suffix, GPT-4o ranks candidates and optimizes later suffixes, and each suffix also guides the rewriting of the seven reviews. In a final live-feedback phase, rankings observed from ChatGPT for CORE's non-winning products are fed back to its optimizer for further rounds. All texts are then frozen; the rankings used as optimization feedback are not counted as evaluation results.

\textbf{Evaluators.} We compare three GEs (ChatGPT with GPT-6, Gemini 3.6 Thinking, and Grok in Expert mode) with five direct-text evaluators (four APIs: GPT-4o, DeepSeek-V4-Pro, Kimi K2.6, and Claude Sonnet 5.5; one local model: Qwen2.5-3B-Instruct \citep{qwen2025qwen25technicalreport}). The GEs receive the published product URLs or a brand URL and rank the six products after reading the pages. The direct-text evaluators receive the frozen descriptions and reviews as text, without web access. The Claude API returned complete rankings for 78 of the 80 brands and refused the remaining two requests, possibly because their content triggered Anthropic's usage policy \citep{anthropic2025usagepolicy}; Claude results therefore use $n=78$.

\textbf{Experimental conditions.} We design experiments on the URL entry point (product URLs or a brand URL), candidate order, public identity, and price, while the product records stay fixed. In the product-URL condition, we shuffle each brand's six links once locally and keep that order for repeated measurements; in the brand-URL condition, the GE opens the brand page, whose product order the site randomizes on each request. For order sensitivity, we use two fixed product-URL sequences on ChatGPT, Gemini, and Grok under equal prices, ordered by the strategies' mean ranks in the initial ChatGPT equal-price run (Table~\ref{tab:ranks}): Original-first (Original, AgenticGEO, C-SEO, Statistics Addition, E-GEO, CORE) and its exact reverse. For identity sensitivity, we replace all six strategy names, including Original, with neutral aliases assigned at random within each brand and applied to page text, URLs, and navigation (Appendix~\ref{app:masking}). For price sensitivity, we compare equal prices with prices frozen within $\pm0.2\%$ of each brand's base price.

\begin{figure}[t]
\begin{tcolorbox}[colback=gray!5!white, colframe=gray!75!black, boxrule=0.4pt, sharp corners, left=4pt, right=4pt, top=3pt, bottom=3pt]
\footnotesize\raggedright
You are a helpful assistant. The user query is: best \{category\}\\
Use the following six candidate product pages: \textless URL 1\textgreater\ \dots\ \textless URL 6\textgreater\\
Before ranking, open each individual product page and read its product description and all displayed customer comments. Do not judge from the brand-page summaries alone.\\
After reading, recommend the products by producing a ranked list from most to least relevant to the user query. Only use the information provided on these pages.\\
Return a JSON object containing: (1) ``ranking'': the six exact product names from rank 1 (best) to rank 6 (worst); (2) ``reading\_log'': for each product, ``description\_read'' (full, partial, or none), ``comments\_displayed'', ``comments\_read'', and ``access\_limitation''; (3) ``limitations'': null if all descriptions and displayed comments were fully read, otherwise the actual access gaps in one sentence.
\end{tcolorbox}
\vspace{-6pt}
\caption{Ranking prompt for the product-URL condition, abridged. Brand-URL requests replace the URL list with one brand page; direct-text evaluators receive the same records as a JSON array and return only the ranked list. Full prompts are in Appendix~\ref{app:prompts}.}
\vspace{-15pt}

\Description{The abridged ranking prompt given to the GEs: the query, six product URLs, reading instructions, and the JSON output format.}
\label{fig:prompt}
\end{figure}

\textbf{Protocol and metrics.} All GE conditions use one ranking prompt adapted from CORE \citep{jin2026core}, with small changes for webpage access and output format (Figure~\ref{fig:prompt}; full prompts in Appendix~\ref{app:prompts}). The prompt asks the evaluator to open the six product pages, read the descriptions and displayed reviews, and return a complete ranking with a reading record per product. For each request we keep the content snapshot, URLs, prompt, capture time, model-setting evidence, raw response, and parsed ranking; only complete permutations of the six candidates enter the statistics. We measure each strategy's overall performance by its Top-1 rate and mean rank. To compare two conditions, we compute Spearman's $\rho$ between the two rankings within each brand and average it across brands, and we report the share of brands with the same top product and the same last product, and the mean absolute rank difference. Uncertainty in paired comparisons is given as a 95\% bootstrap interval from 10,000 brand resamples \citep{efron1993bootstrap}. Missing or incomplete rankings are excluded from the denominator.

\subsection{Evaluation}
\label{sec:results}

We first compare rankings across evaluators (\hyperref[rq:consistency]{RQ1}), then examine their sensitivity to repeated evaluation, candidate order, public identity, and prices (\hyperref[rq:sensitivity]{RQ2}--\hyperref[rq:price-url]{RQ3}). Finally, we report numerical-faithfulness flags and examine how placing flagged products below unflagged ones changes the recorded rankings (\hyperref[rq:faithfulness]{RQ4}). Each question asks whether a visibility gain measured under one condition survives on the live Web.

\begin{table*}[t]
\centering
\caption{Top-1 rate (\%) / mean rank of each strategy under every evaluation condition, with the winning margin and rank spread of each row, and JEV advisory flags per strategy; bold marks the highest Top-1 rate in each row.}
\label{tab:ranks}
\label{tab:ablations80}
\label{tab:price80}
\label{tab:ranks-flag-adjusted80}
\label{tab:jev-flags80}
\small
\setlength{\tabcolsep}{2pt}
\renewcommand{\arraystretch}{0.92}
\vspace{-5pt}
\begin{tabular*}{\textwidth}{@{\extracolsep{\fill}}lrrrrrr@{\hspace{2.5em}}rr@{}}
\toprule
 & \multicolumn{6}{c}{Top-1 rate (\%) / mean rank per strategy} & \multicolumn{2}{c}{Row summary} \\
\cmidrule(lr){2-7}\cmidrule(l){8-9}
Condition & Original & CORE & C-SEO & E-GEO & AgenticGEO & Statistics Add. & Margin & Spread \\
\midrule
\multicolumn{9}{@{}l}{\textit{ChatGPT (product URLs unless noted)}} \\
$\pm0.2\%$ & 2.5 / 5.28 & \textbf{76.3 / 1.74} & 2.5 / 4.05 & 5.0 / 2.96 & 3.8 / 3.80 & 10.0 / 3.18 & 66.3 & 3.54 \\
Equal, random order & 0.0 / 5.86 & \textbf{82.5 / 1.46} & 0.0 / 4.09 & 10.0 / 2.43 & 0.0 / 4.09 & 7.5 / 3.08 & 72.5 & 4.40 \\
Equal, repeat & 0.0 / 5.91 & \textbf{83.8 / 1.41} & 0.0 / 4.03 & 3.8 / 2.46 & 0.0 / 4.13 & 12.5 / 3.06 & 71.3 & 4.50 \\
Equal, Original-first order & 0.0 / 5.90 & \textbf{73.8 / 1.95} & 0.0 / 3.78 & 13.8 / 2.45 & 0.0 / 4.03 & 12.5 / 2.90 & 60.0 & 3.95 \\
Equal, reverse order & 0.0 / 5.80 & \textbf{93.8 / 1.16} & 0.0 / 4.06 & 3.8 / 2.36 & 0.0 / 4.24 & 2.5 / 3.38 & 90.0 & 4.64 \\
Equal, masked identity & 0.0 / 5.75 & \textbf{76.3 / 1.58} & 0.0 / 4.16 & 2.5 / 2.94 & 0.0 / 4.11 & 21.3 / 2.46 & 55.0 & 4.18 \\
Equal, flag-adjusted & 0.0 / 4.81 & \textbf{86.3 / 1.39} & 0.0 / 3.34 & 13.8 / 2.09 & 0.0 / 3.41 & 0.0 / 5.96 & 72.5 & 4.57 \\
Brand URL, equal & 13.8 / 4.85 & \textbf{53.8 / 2.69} & 0.0 / 3.70 & 26.3 / 2.28 & 1.3 / 4.00 & 5.0 / 3.49 & 27.5 & 2.58 \\
Brand URL, $\pm0.2\%$ & 12.5 / 4.61 & \textbf{46.3 / 2.45} & 7.5 / 3.73 & 8.8 / 3.38 & 11.3 / 3.68 & 13.8 / 3.16 & 32.5 & 2.16 \\
\midrule
\multicolumn{9}{@{}l}{\textit{Grok (product URLs unless noted)}} \\
$\pm0.2\%$ & 0.0 / 5.84 & \textbf{67.5 / 2.21} & 8.8 / 3.00 & 7.5 / 3.01 & 5.0 / 3.63 & 11.3 / 3.31 & 56.2 & 3.63 \\
Equal & 0.0 / 5.91 & \textbf{65.0 / 2.40} & 10.0 / 2.73 & 7.5 / 3.23 & 6.3 / 3.76 & 11.3 / 2.98 & 53.7 & 3.51 \\
Equal, Original-first order & 1.3 / 5.40 & \textbf{37.5 / 3.80} & 11.3 / 2.79 & 26.3 / 2.96 & 2.5 / 3.40 & 21.3 / 2.65 & 11.2 & 2.75 \\
Equal, reverse order & 1.3 / 5.88 & \textbf{50.0 / 2.90} & 12.5 / 2.69 & 6.3 / 3.36 & 5.0 / 3.65 & 25.0 / 2.53 & 25.0 & 3.35 \\
Brand URL, $\pm0.2\%$ & 1.3 / 5.81 & \textbf{71.3 / 2.04} & 3.8 / 3.09 & 8.8 / 2.74 & 2.5 / 3.83 & 12.5 / 3.50 & 58.8 & 3.78 \\
Brand URL, equal & 0.0 / 5.93 & \textbf{70.0 / 2.05} & 7.5 / 2.79 & 12.5 / 2.81 & 0.0 / 4.00 & 10.0 / 3.43 & 57.5 & 3.88 \\
Brand URL, equal, flag-adjusted & 0.0 / 4.85 & \textbf{71.3 / 1.81} & 8.8 / 2.43 & 16.3 / 2.53 & 3.8 / 3.44 & 0.0 / 5.95 & 55.0 & 4.14 \\
\midrule
\multicolumn{9}{@{}l}{\textit{Gemini (product URLs)}} \\
$\pm0.2\%$ & 10.0 / 4.88 & \textbf{66.3 / 2.01} & 2.5 / 3.46 & 7.5 / 3.25 & 8.8 / 3.58 & 5.0 / 3.83 & 56.3 & 2.87 \\
Equal & 11.3 / 4.99 & \textbf{82.5 / 1.39} & 1.3 / 3.44 & 5.0 / 3.16 & 0.0 / 3.90 & 0.0 / 4.13 & 71.2 & 3.60 \\
Equal, Original-first order & 11.3 / 4.49 & 23.8 / 3.85 & 10.0 / 3.34 & \textbf{30.0 / 2.98} & 15.0 / 2.86 & 10.0 / 3.49 & 6.2 & 1.63 \\
Equal, reverse order & 7.5 / 4.85 & \textbf{28.8 / 3.16} & 15.0 / 3.35 & 27.5 / 2.61 & 12.5 / 3.38 & 8.8 / 3.65 & 1.3 & 2.24 \\
\midrule
\multicolumn{9}{@{}l}{\textit{GPT-4o API}} \\
$\pm0.2\%$ & 1.3 / 5.00 & \textbf{57.5 / 2.25} & 11.3 / 3.60 & 8.8 / 3.65 & 6.3 / 3.48 & 15.0 / 3.03 & 42.5 & 2.75 \\
Equal, flag-adjusted & 0.0 / 4.23 & \textbf{66.3 / 1.93} & 11.3 / 2.99 & 12.5 / 2.96 & 10.0 / 2.95 & 0.0 / 5.95 & 53.8 & 4.02 \\
\midrule
\multicolumn{9}{@{}l}{\textit{DeepSeek API}} \\
$\pm0.2\%$ & 2.5 / 5.36 & \textbf{82.5 / 1.34} & 5.0 / 3.43 & 2.5 / 3.55 & 3.8 / 3.66 & 3.8 / 3.66 & 77.5 & 4.02 \\
Equal & 2.5 / 5.36 & \textbf{85.0 / 1.28} & 3.8 / 3.53 & 2.5 / 3.49 & 3.8 / 3.54 & 2.5 / 3.81 & 81.2 & 4.08 \\
Equal, flag-adjusted & 2.5 / 4.48 & \textbf{83.8 / 1.28} & 3.8 / 3.08 & 3.8 / 3.10 & 6.3 / 3.11 & 0.0 / 5.96 & 77.5 & 4.68 \\
\midrule
\multicolumn{9}{@{}l}{\textit{Kimi API}} \\
$\pm0.2\%$ & 3.8 / 5.19 & \textbf{56.3 / 2.64} & 5.0 / 3.09 & 8.8 / 2.79 & 2.5 / 4.25 & 23.8 / 3.05 & 32.5 & 2.55 \\
Equal & 0.0 / 5.64 & \textbf{55.0 / 2.64} & 8.8 / 2.93 & 12.5 / 2.44 & 0.0 / 4.30 & 23.8 / 3.06 & 31.2 & 3.20 \\
Equal, flag-adjusted & 2.5 / 4.63 & \textbf{58.8 / 2.25} & 12.5 / 2.45 & 21.3 / 2.14 & 5.0 / 3.59 & 0.0 / 5.95 & 37.5 & 3.81 \\
\midrule
\multicolumn{9}{@{}l}{\textit{Claude API ($n=78$)}} \\
$\pm0.2\%$ & \textbf{44.9 / 1.85} & 2.6 / 5.56 & 7.7 / 3.04 & 41.0 / 1.91 & 2.6 / 3.90 & 1.3 / 4.74 & 3.9 & 3.71 \\
Equal & \textbf{53.8 / 1.59} & 0.0 / 5.87 & 0.0 / 3.13 & 44.9 / 1.59 & 0.0 / 4.05 & 1.3 / 4.77 & 8.9 & 4.28 \\
Equal, flag-adjusted & \textbf{55.1 / 1.55} & 0.0 / 4.91 & 0.0 / 3.03 & 44.9 / 1.67 & 0.0 / 3.88 & 0.0 / 5.96 & 10.2 & 4.41 \\
\midrule
\multicolumn{9}{@{}l}{\textit{Qwen2.5-3B-Instruct (local)}} \\
$\pm0.2\%$ & 7.5 / 4.45 & \textbf{23.8 / 2.96} & 20.0 / 3.15 & 22.5 / 3.34 & 13.8 / 3.70 & 12.5 / 3.40 & 1.3 & 1.49 \\
Equal, flag-adjusted & 5.0 / 3.79 & \textbf{32.5 / 2.43} & 22.5 / 2.70 & 26.3 / 2.80 & 13.8 / 3.33 & 0.0 / 5.96 & 6.2 & 3.53 \\
\midrule
\multicolumn{9}{@{}l}{\textit{JEV advisory flags (86 of 400 rewritten descriptions)}} \\
Flagged descriptions, of 80 & -- & 1 & 0 & 2 & 3 & 80 & -- & -- \\
\bottomrule
\end{tabular*}
\par\vspace{2pt}\parbox{\textwidth}{\footnotesize  Equal and $\pm0.2\%$ denote the price setting, and a lower mean rank is better. Margin is the best minus the second-best Top-1 rate in the row (percentage points). Spread is the worst minus the best mean rank. Flag-adjusted rows move JEV-flagged products below unflagged ones and keep the order within each group. Paired intervals are in Table~\ref{tab:paired80}.}
\end{table*}

\begin{table*}[t]
\centering
\caption{Mean-rank change between conditions with 95\% brand-bootstrap intervals ($n=80$ per contrast; negative favors the first condition).}
\label{tab:paired80}
\small
\setlength{\tabcolsep}{3pt}
\vspace{-3pt}
\begin{tabular*}{\textwidth}{@{\extracolsep{\fill}}lrrrrrr@{}}
\toprule
 & \multicolumn{3}{c}{Candidate order and public identity (ChatGPT, equal prices)} & \multicolumn{3}{c}{$\pm0.2\%$ minus equal prices (product URLs)} \\
\cmidrule(lr){2-4}\cmidrule(lr){5-7}
Method & Reverse vs.\ Original-first & Masked vs.\ repeat & Repeat vs.\ initial & ChatGPT & Grok & Gemini \\
\midrule
Original & -0.10 {\scriptsize[-0.29, 0.06]} & -0.16 {\scriptsize[-0.29, -0.05]} & +0.05 {\scriptsize[-0.04, 0.15]} & -0.59 {\scriptsize[-0.84, -0.38]} & -0.08 {\scriptsize[-0.18, 0.03]} & -0.11 {\scriptsize[-0.65, 0.40]} \\
CORE & -0.79 {\scriptsize[-1.19, -0.41]} & +0.16 {\scriptsize[-0.06, 0.40]} & -0.05 {\scriptsize[-0.31, 0.21]} & +0.28 {\scriptsize[-0.06, 0.64]} & -0.19 {\scriptsize[-0.78, 0.41]} & +0.63 {\scriptsize[0.21, 1.06]} \\
C-SEO & +0.29 {\scriptsize[0.06, 0.51]} & +0.14 {\scriptsize[-0.08, 0.35]} & -0.06 {\scriptsize[-0.24, 0.13]} & -0.04 {\scriptsize[-0.35, 0.28]} & +0.28 {\scriptsize[-0.06, 0.61]} & +0.03 {\scriptsize[-0.29, 0.34]} \\
E-GEO & -0.09 {\scriptsize[-0.29, 0.11]} & +0.48 {\scriptsize[0.24, 0.73]} & +0.04 {\scriptsize[-0.13, 0.21]} & +0.54 {\scriptsize[0.26, 0.84]} & -0.21 {\scriptsize[-0.55, 0.13]} & +0.09 {\scriptsize[-0.33, 0.51]} \\
AgenticGEO & +0.21 {\scriptsize[0.00, 0.43]} & -0.01 {\scriptsize[-0.23, 0.20]} & +0.04 {\scriptsize[-0.15, 0.23]} & -0.29 {\scriptsize[-0.60, 0.03]} & -0.14 {\scriptsize[-0.46, 0.19]} & -0.33 {\scriptsize[-0.64, -0.01]} \\
Statistics Addition & +0.48 {\scriptsize[0.21, 0.73]} & -0.60 {\scriptsize[-0.91, -0.29]} & -0.01 {\scriptsize[-0.29, 0.25]} & +0.10 {\scriptsize[-0.23, 0.43]} & +0.34 {\scriptsize[-0.05, 0.71]} & -0.30 {\scriptsize[-0.66, 0.05]} \\
\bottomrule
\end{tabular*}
\par\vspace{2pt}\parbox{\textwidth}{\footnotesize Intervals use 10,000 brand resamples without multiplicity adjustment.}
\end{table*}

\begin{table}[!t]
\centering
\caption{Ranking agreement between evaluators under frozen $\pm0.2\%$ prices and between ChatGPT presentation conditions under equal prices.}
\vspace{-4pt}

\label{tab:agreement}
\label{tab:agreement-supp80}
\small
\setlength{\tabcolsep}{2pt}
\renewcommand{\arraystretch}{0.92}
\vspace{-3pt}
\begin{tabular*}{\columnwidth}{@{\extracolsep{\fill}}lrrrr@{}}
\toprule
Pair & $\rho$ [95\% CI] & Same Top-1 & Same last & Gap \\
\midrule
\multicolumn{5}{@{}l}{\textit{GEs}} \\
ChatGPT / Grok & 0.48 [0.39, 0.58] & 66.3\% & 60.0\% & 1.14 \\
ChatGPT / Gemini & 0.37 [0.26, 0.48] & 57.5\% & 38.8\% & 1.35 \\
Grok / Gemini & 0.32 [0.21, 0.43] & 48.8\% & 47.5\% & 1.38 \\
\midrule
\multicolumn{5}{@{}l}{\textit{Direct-text evaluators vs.\ GEs}} \\
GPT-4o API / ChatGPT & 0.36 [0.25, 0.46] & 55.0\% & 45.0\% & 1.37 \\
GPT-4o API / Grok & 0.32 [0.23, 0.41] & 48.8\% & 52.5\% & 1.45 \\
GPT-4o API / Gemini & 0.28 [0.17, 0.38] & 46.3\% & 38.8\% & 1.54 \\
DeepSeek / ChatGPT & 0.50 [0.41, 0.59] & 66.3\% & 63.8\% & 1.15 \\
DeepSeek / Grok & 0.44 [0.35, 0.52] & 58.8\% & 68.8\% & 1.27 \\
DeepSeek / Gemini & 0.32 [0.21, 0.42] & 57.5\% & 38.8\% & 1.47 \\
Kimi / ChatGPT & 0.38 [0.29, 0.48] & 53.8\% & 50.0\% & 1.33 \\
Kimi / Grok & 0.41 [0.31, 0.51] & 51.3\% & 61.3\% & 1.31 \\
Kimi / Gemini & 0.25 [0.14, 0.35] & 41.3\% & 41.3\% & 1.56 \\
Claude / ChatGPT & -0.22 [-0.33, -0.12] & 9.0\% & 12.8\% & 2.09 \\
Claude / Grok & -0.31 [-0.41, -0.22] & 5.1\% & 10.3\% & 2.19 \\
Claude / Gemini & -0.26 [-0.35, -0.17] & 6.4\% & 7.7\% & 2.13 \\
Qwen2.5-3B / ChatGPT & 0.13 [0.04, 0.22] & 12.5\% & 21.3\% & 1.80 \\
Qwen2.5-3B / Grok & 0.20 [0.11, 0.30] & 23.8\% & 27.5\% & 1.69 \\
Qwen2.5-3B / Gemini & 0.17 [0.07, 0.27] & 25.0\% & 22.5\% & 1.72 \\
\midrule
\multicolumn{5}{@{}l}{\textit{ChatGPT presentation conditions}} \\
Repeat / initial & 0.85 [0.80, 0.90] & 80.0\% & 90.0\% & 0.49 \\
Opposite fixed orders & 0.74 [0.68, 0.80] & 75.0\% & 78.8\% & 0.71 \\
Masked / repeat & 0.79 [0.75, 0.84] & 78.8\% & 83.8\% & 0.65 \\
\bottomrule
\end{tabular*}
\par\vspace{2pt}\parbox{\columnwidth}{\footnotesize Claude pairs use $n=78$, others $n=80$; intervals from 10,000 paired brand resamples. Gap: mean absolute rank difference.}
\vspace{-8pt}
\end{table}

\begin{table}[!t]
\centering
\vspace{-4pt}
\caption{Direct-text API sensitivity to reversing each brand's saved candidate order (equal and $\pm0.2\%$ prices).}
\label{tab:api-order80}
\small
\setlength{\tabcolsep}{3pt}
\vspace{-10pt}
\begin{tabular*}{\columnwidth}{@{\extracolsep{\fill}}llrrr@{}}
\toprule
Evaluator & Price & $n$ & $\rho$ [95\% CI] & Same Top-1 \\
\midrule
DeepSeek & Equal & 80 & 0.33 [0.24, 0.43] & 72.5\% \\
DeepSeek & $\pm0.2\%$ & 80 & 0.37 [0.27, 0.47] & 65.0\% \\
Kimi & Equal & 80 & 0.49 [0.40, 0.57] & 51.3\% \\
Kimi & $\pm0.2\%$ & 80 & 0.39 [0.29, 0.50] & 52.5\% \\
Claude & Equal & 78 & 0.85 [0.82, 0.89] & 43.6\% \\
Claude & $\pm0.2\%$ & 78 & 0.76 [0.69, 0.81] & 56.4\% \\
\bottomrule
\end{tabular*}
\par\vspace{2pt}\parbox{\columnwidth}{\footnotesize Claude excludes two refused brands; intervals from 10,000 paired brand resamples.}
\vspace{13pt}

\end{table}

\begin{table}[!t]
\centering
\caption{JEV accuracy on curated numerical test subsets.}
\vspace{-4pt}

\label{tab:jev-benchmark}
\small
\setlength{\tabcolsep}{3pt}
\vspace{-5pt}
\begin{tabular*}{\columnwidth}{@{\extracolsep{\fill}}lrrrr@{}}
\toprule
Source & $n$ & Accuracy (\%) & FN & FP \\
\midrule
SciTab~\citep{lu-etal-2023-scitab} & 24 & 91.67 & 1 & 1 \\
TabFact~\citep{Chen2020TabFact:} & 85 & 92.94 & 4 & 2 \\
FEVEROUS~\citep{aly2021feverous} & 85 & 88.24 & 8 & 2 \\
\midrule
Overall & 194 & 90.72 & 13 & 5 \\
\bottomrule
\end{tabular*}
\par\vspace{2pt}\parbox{\columnwidth}{\footnotesize 194 held-out test cases from assistant-screened numerical subsets of the three datasets. FN/FP: missed/false flags.}
\vspace{-20pt}
\end{table}

\begin{table}[!t]
\centering
\vspace{8pt}
\caption{JEV agreement with three human reviewers on 50 sampled descriptions.}
\vspace{-5pt}
\label{tab:human50}
\small
\setlength{\tabcolsep}{3pt}
\vspace{-3pt}
\begin{tabular*}{\columnwidth}{@{\extracolsep{\fill}}lrrr@{}}
\toprule
Reviewer & Flag & No flag & Agrees with JEV \\
\midrule
A & 10 & 40 & 50/50 \\
B & 10 & 40 & 50/50 \\
C & 11 & 39 & 49/50 \\
\midrule
Two-of-three majority & 10 & 40 & 50/50 \\
\bottomrule
\end{tabular*}
\par\vspace{2pt}\parbox{\columnwidth}{\footnotesize Ten descriptions per strategy from 36 brands; 49/50 unanimous, Fleiss' $\kappa=0.96$. A fourth reviewer was excluded post hoc; including them leaves all majorities unchanged ($\kappa=0.53$).}
\vspace{-18pt}
\end{table}

\textbf{Ranking consistency across evaluators (RQ1).} Under frozen $\pm0.2\%$ prices, CORE leads Top-1 rates across seven evaluators (Table~\ref{tab:ranks}). On the three GEs, CORE's Top-1 rate is 76.3\% (ChatGPT), 67.5\% (Grok), and 66.3\% (Gemini), with mean ranks of 1.74, 2.21, and 2.01 and margins of 56 to 66 percentage points over the second-best strategy. The exception is the Claude API, which ranks Original first (44.9\%) and E-GEO second (41.0\%), while CORE comes last with a mean rank of 5.56. Qwen also ranks CORE first, but with a Top-1 rate of 23.8\% and a spread of 1.49 it barely separates the strategies.

\begin{figure}[!htb]
\vspace{-12pt}

  \centering
  \includegraphics[width=\columnwidth]{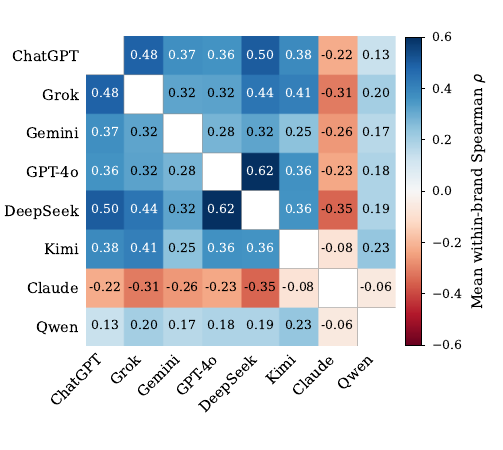}
  \vspace{-30pt}
  \caption{Mean within-brand Spearman $\rho$ across evaluators, with frozen $\pm0.2\%$ prices and product URLs for GEs ($n=80$; $n=78$ with Claude).}
  \Description{An eight by eight heatmap of rank correlations between ChatGPT, Grok, Gemini, GPT-4o, DeepSeek, Kimi, Claude, and Qwen; Claude's row is negative throughout.}
  \vspace{-10pt}
  \label{fig:agreement-heatmap}
  
\end{figure}

Agreement in the aggregate does not imply agreement on individual brands (Table~\ref{tab:agreement}). Each pair of GEs selects the same top product for only 48.8\% to 66.3\% of brands, with mean $\rho$ between 0.32 and 0.48; for the active-layer thaw-bed planter, for example, Grok selects Statistics Addition and Gemini selects CORE. The direct-text evaluators differ widely in how closely they follow the GEs (Figure~\ref{fig:agreement-heatmap}): DeepSeek reaches $\rho=0.50$ and the same top product as ChatGPT on 66.3\% of brands, matching the agreement among the GEs themselves; GPT-4o and Kimi are intermediate ($\rho$ from 0.25 to 0.41); Qwen is weak ($\rho\leq0.20$); and Claude correlates negatively with all three GEs ($\rho$ from $-0.31$ to $-0.22$, same top product below 10\%). Agreement on the last product is similarly limited, ranging from 7.7\% (Claude and Gemini) to 68.8\% (DeepSeek and Grok), and it exceeds agreement on the top product in only half of the pairs.

\begin{tcolorbox}[colback=gray!5!white, colframe=gray!75!black, boxrule=0.4pt, sharp corners, breakable, lines before break=4]
\textbf{RQ1. How consistent is GEO visibility across GEs and direct-text evaluators?}

\textbf{Answer.} Visibility is consistent in the aggregate and inconsistent in detail: seven of the eight evaluators rank CORE first (Table~\ref{tab:ranks}). The three GEs reach a mean within-brand Spearman $\rho$ of only 0.32 to 0.48 and pick the same top product for 48.8\% to 66.3\% of brands (Table~\ref{tab:agreement}). Against the same GEs the direct-text evaluators range from $\rho=0.50$ for DeepSeek to $\rho=-0.31$ for Claude. Only the aggregate part of a gain measured by direct-text evaluation survives on the live Web.
\end{tcolorbox}

\textbf{Repeatability and presentation sensitivity (RQ2).}
Repeating the ChatGPT evaluation with the same text, prompt, and product-URL order gives stable aggregate results: the two runs select the same top product for 80.0\% of brands and the same last product for 90.0\%, with mean $\rho=0.85$, and CORE's Top-1 rate is 82.5\% and 83.8\% (Tables~\ref{tab:ranks} and~\ref{tab:agreement}); no strategy's mean-rank change between the two runs has an interval that excludes zero (Table~\ref{tab:paired80}).

Candidate order changes the observed ranking. On ChatGPT, CORE's Top-1 rate under the two fixed orders is 73.8\% (Original-first) and 93.8\% (reverse), a difference of 20 percentage points, and its margin grows from 60.0 to 90.0 (Table~\ref{tab:ranks}). In the paired contrasts (Table~\ref{tab:paired80}), the reverse order improves CORE's mean rank by 0.79 and worsens C-SEO's and Statistics Addition's by 0.29 and 0.48, all with intervals excluding zero. On Grok, the same two sequences give CORE Top-1 rates of 37.5\% and 50.0\%; on Gemini, the highest Top-1 rate moves from E-GEO (30.0\%) under Original-first to CORE (28.8\%) under the reverse order (Table~\ref{tab:ranks}). ChatGPT's random-order results fall between its two fixed orders, which is why LIVE-GEO randomizes the candidate order for each brand and keeps that order for repeated measurements. The direct-text APIs are also order-sensitive (Table~\ref{tab:api-order80}): after reversing the saved order, $\rho$ is 0.33 to 0.37 for DeepSeek, 0.39 to 0.49 for Kimi, and 0.76 to 0.85 for Claude, with the same top product on 43.6\% to 72.5\% of brands.

Compared with the unmasked repeat, the masked run (Appendix~\ref{app:masking}) gives CORE a lower Top-1 rate (76.3\% versus 83.8\%) and Statistics Addition a higher rate (21.3\% versus 12.5\%) (Table~\ref{tab:ranks}); Statistics Addition's mean rank improves by 0.60 with an interval excluding zero (Table~\ref{tab:paired80}). The masked and unmasked runs select the same top product for 78.8\% of brands, with $\rho=0.79$ (Table~\ref{tab:agreement}).

\begin{tcolorbox}[colback=gray!5!white, colframe=gray!75!black, boxrule=0.4pt, sharp corners, breakable, lines before break=4]

\textbf{RQ2. How sensitive is visibility to repeated evaluation, candidate order, and public identity?}

\textbf{Answer.} Visibility is insensitive to repetition and sensitive to presentation. A repeat under the same presentation gives $\rho=0.85$ and the same top product for 80.0\% of brands (Table~\ref{tab:agreement}). CORE's Top-1 rate on ChatGPT ranges from 73.8\% to 93.8\% across opposite orders and is 83.8\% in the unmasked repeat versus 76.3\% in the masked run (Table~\ref{tab:ranks}). A gain that depends on one presentation does not survive another.

\end{tcolorbox}
\vspace{-5pt}
\textbf{Price and URL entry conditions (RQ3).}
With product URLs, the leading strategy stays the same between equal and frozen $\pm0.2\%$ prices, while the values change (Table~\ref{tab:ranks}). With product URLs, CORE's Top-1 rate moves from 82.5\% to 76.3\% on ChatGPT, from 65.0\% to 67.5\% on Grok, and from 82.5\% to 66.3\% on Gemini, and it remains first on all three GEs. In the paired comparison (Table~\ref{tab:paired80}), only four intervals exclude zero: on ChatGPT, Original's mean rank improves by 0.59 and E-GEO's worsens by 0.54; on Gemini, CORE's worsens by 0.63 and AgenticGEO's improves by 0.33.

Rankings also differ across URL-entry conditions. With $\pm0.2\%$ prices, CORE's Top-1 rate on Grok is 71.3\% with the brand URL and 67.5\% with product URLs, with mean ranks of 2.04 and 2.21; on ChatGPT the values are 46.3\% and 76.3\%, with mean ranks of 2.45 and 1.74 (Table~\ref{tab:ranks}).

\vspace{-4pt}
\begin{tcolorbox}[colback=gray!5!white, colframe=gray!75!black, boxrule=0.4pt, sharp corners, breakable, lines before break=4]
\textbf{RQ3. How do small price differences and URL entry conditions affect observed visibility?}

\textbf{Answer.} Values change, but the Top-1 leader remains. CORE's Top-1 rate differs by at most 16.2 percentage points between equal and $\pm0.2\%$ prices (82.5\% versus 66.3\% on Gemini) and 30.0 points between product URLs and the brand page (76.3\% versus 46.3\% on ChatGPT) (Table~\ref{tab:ranks}). CORE leads Top-1 rates throughout, but E-GEO leads mean rank for ChatGPT's equal-price brand-URL condition.
\end{tcolorbox}

\textbf{Numerical-faithfulness flags (RQ4).}
JEV flags 86 of the 400 rewritten descriptions (Table~\ref{tab:ranks}, last row): all 80 Statistics Addition descriptions, 3 from AgenticGEO, 2 from E-GEO, 1 from CORE, and none from C-SEO. The Statistics Addition prompt asks for compelling statistics and permits hypothetical data, so numbers unsupported by the specifications are common, for example an ``18\% improvement in heat retention''; flagged descriptions from other strategies also contain unsupported comparisons, such as CORE's ``30\% energy savings'' and AgenticGEO's ``up to 60\% less water use.''

To test the verifier, we apply it to 194 numerical test cases drawn from SciTab, TabFact, and FEVEROUS, where JEV agrees with the reference labels on 176 cases (90.72\%; Table~\ref{tab:jev-benchmark}). We also validate it against human reviewers: three researchers independently reviewed 50 randomly sampled descriptions (ten per strategy) without seeing JEV's output. JEV agrees with the two-of-three majority on all 50 descriptions, 49 of which are unanimous, and Fleiss' $\kappa$ is 0.96 (Table~\ref{tab:human50}); all ten majority flags belong to Statistics Addition.

We place flagged products after unflagged ones and keep the Original order within each group, which gives the flag-adjusted rows in Table~\ref{tab:ranks}. After the adjustment, Statistics Addition has a Top-1 rate of zero in every condition and a mean rank of 5.95 to 5.96; under the ChatGPT product-URL condition, CORE's Top-1 rate rises from 82.5\% to 86.3\% and E-GEO's from 10.0\% to 13.8\%.
\vspace{-5pt}
\begin{tcolorbox}[colback=gray!5!white, colframe=gray!75!black, boxrule=0.4pt, sharp corners, breakable, lines before break=4]
\textbf{RQ4. How reliable are numerical-faithfulness flags, and how does applying them affect visibility?}

\textbf{Answer.} The flags are reliable and they remove only gains built on unsupported numbers. JEV agrees with the reference labels on 90.72\% of 194 benchmark cases (Table~\ref{tab:jev-benchmark}) and with the human majority on all 50 reviewed descriptions (Table~\ref{tab:human50}). Moving flagged products below unflagged ones sets Statistics Addition's Top-1 rate to zero in every condition and raises CORE's from 82.5\% to 86.3\% on ChatGPT (Table~\ref{tab:ranks}). Gains built on numbers the specifications do not support do not survive the flags.
\end{tcolorbox}
\vspace{-12pt}
\subsection{Results and Discussion}
\label{sec:discussion}

\textbf{Do GEO gains survive the live Web?}
Some gains survive and others do not. On the same frozen text and under the randomized candidate order, four direct-text evaluators (GPT-4o, DeepSeek, Kimi, and Qwen) and all three GEs give CORE the highest Top-1 rate, and this Top-1 lead holds under both price settings and both entry points on ChatGPT and Grok (RQ1, RQ3). Two things do not carry over. The ordering measured under Claude reverses on the GEs, and agreement in the aggregate does not extend to brands: the GEs pick different winners for a third to a half of the brands, the brand-level agreement between direct-text evaluators and GEs is lower still, and only DeepSeek approaches the level seen among the GEs. Part of the aggregate gain therefore survives, whereas much of the brand-level gain does not.

\textbf{LIVE-GEO as a benchmark.}
The experiments above also test LIVE-GEO itself. It is a public, Kaggle-style platform: fictional products for 80 brands are published on the live Web; users download the original descriptions and reviews, rewrite them with their own method, and upload the result; and each submission becomes a rewritten product of that brand, ranked by GEs under the category query together with the Original and the earlier submissions. Every control used in the experiments is a fixed design of the platform: content snapshots are frozen, so repeated measurements can be reproduced; the candidate order is drawn at random per brand and kept across repeats, so order effects are measured rather than hidden; the public strategy labels are a known treatment factor; prices are perturbed within a fixed band; and JEV flags with human review stop submissions that trade unsupported numbers for rank. The three GEs returned complete rankings for all 80 brands under product URLs, which shows that the benchmark is feasible. Users choose the GEs that can reach the site and their own prompts, run the measurements, and report them themselves; because everyone faces the same frozen set of control products, numbers from different users are comparable. The more submissions arrive, the larger the set of strategies each new submission competes against.

\section{Conclusion}
\label{sec:conclusion}

We present LIVE-GEO, a benchmark that publishes GEO-rewritten products on the live Web and lets GEs rank them under their own retrieval conditions; it freezes content snapshots and records the candidate order, public identity, and price setting, so that visibility measurements from different users are comparable. An evaluation on 80 brands, five strategies, and eight evaluators answers the question in the title: part of the aggregate GEO gain survives on the live Web, but most brand-level gains do not, and LIVE-GEO narrows this gap by letting users measure their own gains where GEs read the pages.

\newpage
\bibliographystyle{ACM-Reference-Format}
\bibliography{references}

@inproceedings{lewis2020rag,
 author = {Lewis, Patrick and Perez, Ethan and Piktus, Aleksandra and Petroni, Fabio and Karpukhin, Vladimir and Goyal, Naman and K\"{u}ttler, Heinrich and Lewis, Mike and Yih, Wen-tau and Rockt\"{a}schel, Tim and Riedel, Sebastian and Kiela, Douwe},
 booktitle = {Advances in Neural Information Processing Systems},
 editor = {H. Larochelle and M. Ranzato and R. Hadsell and M.F. Balcan and H. Lin},
 pages = {9459--9474},
 publisher = {Curran Associates, Inc.},
 title = {Retrieval-Augmented Generation for Knowledge-Intensive {NLP} Tasks},
 url = {https://proceedings.neurips.cc/paper_files/paper/2020/file/6b493230205f780e1bc26945df7481e5-Paper.pdf},
 volume = {33},
 year = {2020}
}

@inproceedings{aggarwal2024geo,
author = {Aggarwal, Pranjal and Murahari, Vishvak and Rajpurohit, Tanmay and Kalyan, Ashwin and Narasimhan, Karthik and Deshpande, Ameet},
title = {{GEO}: Generative Engine Optimization},
year = {2024},
isbn = {9798400704901},
publisher = {Association for Computing Machinery},
address = {New York, NY, USA},
url = {https://doi.org/10.1145/3637528.3671900},
doi = {10.1145/3637528.3671900},
booktitle = {Proceedings of the 30th ACM SIGKDD Conference on Knowledge Discovery and Data Mining},
pages = {5--16},
numpages = {12},
location = {Barcelona, Spain},
series = {KDD '24}
}

@inproceedings{puerto2025cseo,
  title = {C-{SEO} Bench: Does Conversational {SEO} Work?},
  author = {Haritz Puerto and Martin Gubri and Tommaso Green and Seong Joon Oh and Sangdoo Yun},
  booktitle = {Advances in Neural Information Processing Systems},
  volume = {38},
  year = {2025},
  doi = {10.52202/085713-0923},
  url = {https://proceedings.neurips.cc/paper_files/paper/2025/hash/27aa3aeff0f8460a7b43d30fa6c5c032-Abstract-Datasets_and_Benchmarks_Track.html}
}

@misc{bagga2026egeo,
      title={{E-GEO}: A Testbed for Generative Engine Optimization in E-Commerce},
      author={Puneet S. Bagga and Vivek F. Farias and Tamar Korkotashvili and Tianyi Peng and Yuhang Wu},
      year={2026},
      eprint={2511.20867v2},
      archivePrefix={arXiv},
      primaryClass={cs.IR},
      url={https://arxiv.org/abs/2511.20867v2},
}

@inproceedings{kim2026sageo,
author = {Kim, Sunghwan and Jeong, Wooseok and Kim, Serin and Lee, Sangam and Lee, Dongha},
title = {{SAGEO} Arena: A Realistic Environment for Evaluating Search-Augmented Generative Engine Optimization},
year = {2026},
isbn = {9798400722592},
publisher = {Association for Computing Machinery},
address = {New York, NY, USA},
url = {https://doi.org/10.1145/3770855.3818146},
doi = {10.1145/3770855.3818146},
booktitle = {Proceedings of the 32nd ACM SIGKDD Conference on Knowledge Discovery and Data Mining V.2},
pages = {2342--2353},
numpages = {12},
location = {Jeju Island, Republic of Korea},
series = {KDD '26}
}

@misc{zheng2026countergeo,
  title = {Counter-{GEO}-Bench: Evaluating Defenses Against Information-Distorting Generative Engine Optimization},
  author = {Bing Zheng and Zongyao Zhao and Wenming Yang},
  year = {2026},
  eprint = {2609.02316},
  archivePrefix = {arXiv},
  primaryClass = {cs.IR},
  url = {https://arxiv.org/abs/2609.02316}
}

@misc{jin2026core,
  title = {Controlling Output Rankings in Generative Engines for {LLM}-based Search},
  author = {Haibo Jin and Ruoxi Chen and Peiyan Zhang and Yifeng Luo and Huimin Zeng and Man Luo and Haohan Wang},
  year = {2026},
  eprint = {2602.03608v1},
  archivePrefix = {arXiv},
  primaryClass = {cs.CL},
  url = {https://arxiv.org/abs/2602.03608v1}
}

@inproceedings{wu2026autogeo,
title={What Generative Search Engines Like and How to Optimize Web Content Cooperatively},
author={Yujiang Wu and Shanshan Zhong and Yubin Kim and Chenyan Xiong},
booktitle={The Fourteenth International Conference on Learning Representations},
year={2026},
url={https://openreview.net/forum?id=K8EinVWtUB}
}

@inproceedings{tang-etal-2024-minicheck,
    title = "{M}ini{C}heck: Efficient Fact-Checking of {LLM}s on Grounding Documents",
    author = "Tang, Liyan  and
      Laban, Philippe  and
      Durrett, Greg",
    editor = "Al-Onaizan, Yaser  and
      Bansal, Mohit  and
      Chen, Yun-Nung",
    booktitle = "Proceedings of the 2024 Conference on Empirical Methods in Natural Language Processing",
    month = nov,
    year = "2024",
    address = "Miami, Florida, USA",
    publisher = "Association for Computational Linguistics",
    url = "https://aclanthology.org/2024.emnlp-main.499/",
    doi = "10.18653/v1/2024.emnlp-main.499",
    pages = "8818--8847"
}

@inproceedings{kirsten2026characterizingwebsearch,
  title = {Characterizing Web Search in The Age of Generative {AI}},
  author = {Kirsten, Elisabeth and Perdekamp, Jost Gro{\ss}e and Wu, Qinyuan and Upadhyay, Mihir and Gummadi, Krishna P. and Zafar, Muhammad Bilal},
  booktitle = {Findings of the Association for Computational Linguistics: ACL 2026},
  year = {2026},
  address = {San Diego, California, United States},
  publisher = {Association for Computational Linguistics},
  pages = {10827--10848},
  doi = {10.18653/v1/2026.findings-acl.526},
  url = {https://aclanthology.org/2026.findings-acl.526/}
}

@misc{yuan2026agenticgeo,
  title = {{AgenticGEO}: A Self-Evolving Agentic System for Generative Engine Optimization},
  author = {Jiaqi Yuan and Jialu Wang and Zihan Wang and Qingyun Sun and Ruijie Wang and Jianxin Li},
  year = {2026},
  eprint = {2603.20213},
  archivePrefix = {arXiv},
  url = {https://arxiv.org/abs/2603.20213}
}

@inproceedings{dhuliawala-etal-2024-chain,
  title = {Chain-of-Verification Reduces Hallucination in Large Language Models},
  author = {Dhuliawala, Shehzaad and Komeili, Mojtaba and Xu, Jing and Raileanu, Roberta and Li, Xian and Celikyilmaz, Asli and Weston, Jason},
  booktitle = {Findings of the Association for Computational Linguistics: ACL 2024},
  year = {2024},
  pages = {3563--3578},
  publisher = {Association for Computational Linguistics},
  doi = {10.18653/v1/2024.findings-acl.212},
  url = {https://aclanthology.org/2024.findings-acl.212/}
}

@inproceedings{gao-etal-2023-rarr,
  title = {{RARR}: Researching and Revising What Language Models Say, Using Language Models},
  author = {Gao, Luyu and Dai, Zhuyun and Pasupat, Panupong and Chen, Anthony and Chaganty, Arun Tejasvi and Fan, Yicheng and Zhao, Vincent and Lao, Ni and Lee, Hongrae and Juan, Da-Cheng and Guu, Kelvin},
  booktitle = {Proceedings of the 61st Annual Meeting of the Association for Computational Linguistics (Volume 1: Long Papers)},
  year = {2023},
  pages = {16477--16508},
  publisher = {Association for Computational Linguistics},
  doi = {10.18653/v1/2023.acl-long.910},
  url = {https://aclanthology.org/2023.acl-long.910/}
}

@inproceedings{pfrommer-etal-2024-ranking,
    title = "Ranking Manipulation for Conversational Search Engines",
    author = "Pfrommer, Samuel  and
      Bai, Yatong  and
      Gautam, Tanmay  and
      Sojoudi, Somayeh",
    editor = "Al-Onaizan, Yaser  and
      Bansal, Mohit  and
      Chen, Yun-Nung",
    booktitle = "Proceedings of the 2024 Conference on Empirical Methods in Natural Language Processing",
    month = nov,
    year = "2024",
    address = "Miami, Florida, USA",
    publisher = "Association for Computational Linguistics",
    url = "https://aclanthology.org/2024.emnlp-main.534/",
    doi = "10.18653/v1/2024.emnlp-main.534",
    pages = "9523--9552"
}

@misc{kumar2024manipulatingllms,
  title = {Manipulating Large Language Models to Increase Product Visibility},
  author = {Kumar, Aounon and Lakkaraju, Himabindu},
  year = {2024},
  eprint = {2404.07981},
  archivePrefix = {arXiv},
  url = {https://arxiv.org/abs/2404.07981}
}

@inproceedings{nestaas2025adversarialseo,
  title = {Adversarial Search Engine Optimization for Large Language Models},
  author = {Nestaas, Fredrik and Debenedetti, Edoardo and Tram{\`e}r, Florian},
  booktitle = {International Conference on Learning Representations},
  year = {2025},
  url = {https://iclr.cc/virtual/2025/poster/28746}
}

@inproceedings{liu-etal-2023-evaluating,
    title = "Evaluating Verifiability in Generative Search Engines",
    author = "Liu, Nelson  and
      Zhang, Tianyi  and
      Liang, Percy",
    editor = "Bouamor, Houda  and
      Pino, Juan  and
      Bali, Kalika",
    booktitle = "Findings of the {A}ssociation for {C}omputational {L}inguistics: {EMNLP} 2023",
    month = dec,
    year = "2023",
    address = "Singapore",
    publisher = "Association for Computational Linguistics",
    url = "https://aclanthology.org/2023.findings-emnlp.467/",
    doi = "10.18653/v1/2023.findings-emnlp.467",
    pages = "7001--7025"
}

@inproceedings{white2025livebench,
  title = {{LiveBench}: A Challenging, Contamination-Limited {LLM} Benchmark},
  author = {White, Colin and Dooley, Samuel and Roberts, Manley and Pal, Arka and Feuer, Benjamin and Jain, Siddhartha and Shwartz-Ziv, Ravid and Jain, Neel and Saifullah, Khalid and Dey, Sreemanti and Shubh-Agrawal and Sandha, Sandeep Singh and Naidu, Siddartha and Hegde, Chinmay and LeCun, Yann and Goldstein, Tom and Neiswanger, Willie and Goldblum, Micah},
  booktitle = {International Conference on Learning Representations},
  year = {2025},
  url = {https://proceedings.iclr.cc/paper_files/paper/2025/hash/e4a46394ba5378b3f9a186a5b4c650d1-Abstract-Conference.html}
}

@inproceedings{chiang2024chatbotarena,
  title = {Chatbot Arena: An Open Platform for Evaluating {LLMs} by Human Preference},
  author = {Chiang, Wei-Lin and Zheng, Lianmin and Sheng, Ying and Angelopoulos, Anastasios Nikolas and Li, Tianle and Li, Dacheng and Zhu, Banghua and Zhang, Hao and Jordan, Michael and Gonzalez, Joseph E. and Stoica, Ion},
  booktitle = {Proceedings of the 41st International Conference on Machine Learning},
  series = {Proceedings of Machine Learning Research},
  volume = {235},
  pages = {8359--8388},
  year = {2024},
  url = {https://proceedings.mlr.press/v235/chiang24b.html}
}

@book{efron1993bootstrap,
  title = {An Introduction to the Bootstrap},
  author = {Efron, Bradley and Tibshirani, Robert J.},
  publisher = {Chapman \& Hall/CRC},
  year = {1993}
}

@inproceedings{thakur2021beir,
 author = {Thakur, Nandan and Reimers, Nils and R\"{u}ckl\'{e}, Andreas and Srivastava, Abhishek and Gurevych, Iryna},
 booktitle = {Proceedings of the Neural Information Processing Systems Track on Datasets and Benchmarks},
 editor = {J. Vanschoren and S. Yeung},
 pages = {},
 title = {BEIR: A Heterogeneous Benchmark for Zero-shot Evaluation of Information Retrieval Models},
 url = {https://datasets-benchmarks-proceedings.neurips.cc/paper_files/paper/2021/file/65b9eea6e1cc6bb9f0cd2a47751a186f-Paper-round2.pdf},
 volume = {1},
 year = {2021}
}

@inproceedings{nguyen2016msmarco,
  title = {{MS MARCO}: A Human Generated {MA}chine Reading {CO}mprehension Dataset},
  author = {Nguyen, Tri and Rosenberg, Mir and Song, Xia and Gao, Jianfeng and Tiwary, Saurabh and Majumder, Rangan and Deng, Li},
  booktitle = {Proceedings of the Workshop on Cognitive Computation: Integrating Neural and Symbolic Approaches, co-located with NIPS 2016},
  series = {CEUR Workshop Proceedings},
  volume = {1773},
  year = {2016},
  url = {https://ceur-ws.org/Vol-1773/CoCoNIPS_2016_paper9.pdf}
}

@misc{anthropic2025usagepolicy,
  author       = {{Anthropic}},
  title        = {Usage Policy},
  year         = {2026},
  howpublished = {\url{https://www.anthropic.com/legal/aup}},
}

@inproceedings{sainz-etal-2023-nlp,
    title = "{NLP} Evaluation in trouble: On the Need to Measure {LLM} Data Contamination for each Benchmark",
    author = "Sainz, Oscar  and
      Campos, Jon  and
      Garc{\'i}a-Ferrero, Iker  and
      Etxaniz, Julen  and
      de Lacalle, Oier Lopez  and
      Agirre, Eneko",
    editor = "Bouamor, Houda  and
      Pino, Juan  and
      Bali, Kalika",
    booktitle = "Findings of the {A}ssociation for {C}omputational {L}inguistics: {EMNLP} 2023",
    month = dec,
    year = "2023",
    address = "Singapore",
    publisher = "Association for Computational Linguistics",
    url = "https://aclanthology.org/2023.findings-emnlp.722/",
    doi = "10.18653/v1/2023.findings-emnlp.722",
    pages = "10776--10787"
}

@inproceedings{jain2025livecodebench,
  title = {{LiveCodeBench}: Holistic and Contamination Free Evaluation of Large Language Models for Code},
  author = {Jain, Naman and Han, King and Gu, Alex and Li, Wen-Ding and Yan, Fanjia and Zhang, Tianjun and Wang, Sida I. and Solar-Lezama, Armando and Sen, Koushik and Stoica, Ion},
  booktitle = {International Conference on Learning Representations},
  year = {2025},
  url = {https://proceedings.iclr.cc/paper_files/paper/2025/hash/94074dd5a072d28ff75a76dabed43767-Abstract-Conference.html}
}

@misc{nakano2021webgpt,
  title = {{WebGPT}: Browser-assisted question-answering with human feedback},
  author = {Nakano, Reiichiro and Hilton, Jacob and Balaji, Suchir and Wu, Jeff and Ouyang, Long and Kim, Christina and Hesse, Christopher and Jain, Shantanu and Kosaraju, Vineet and Saunders, William and Jiang, Xu and Cobbe, Karl and Eloundou, Tyna and Krueger, Gretchen and Button, Kevin and Knight, Matthew and Chess, Benjamin and Schulman, John},
  year = {2021},
  eprint = {2112.09332},
  archivePrefix = {arXiv},
  url = {https://arxiv.org/abs/2112.09332}
}

@misc{gao2023ragsurvey,
  title = {Retrieval-Augmented Generation for Large Language Models: A Survey},
  author = {Gao, Yunfan and Xiong, Yun and Gao, Xinyu and Jia, Kangxiang and Pan, Jinliu and Bi, Yuxi and Dai, Yi and Sun, Jiawei and Wang, Meng and Wang, Haofen},
  year = {2023},
  eprint = {2312.10997},
  archivePrefix = {arXiv},
  url = {https://arxiv.org/abs/2312.10997}
}

@inproceedings{venkit2025answerengines,
  title = {Search Engines in the {AI} Era: A Qualitative Understanding to the False Promise of Factual and Verifiable Source-Cited Responses in {LLM}-based Search},
  author = {Narayanan Venkit, Pranav and Laban, Philippe and Zhou, Yilun and Mao, Yixin and Wu, Chien-Sheng},
  booktitle = {Proceedings of the 2025 ACM Conference on Fairness, Accountability, and Transparency},
  year = {2025},
  doi = {10.1145/3715275.3732089},
  url = {https://doi.org/10.1145/3715275.3732089}
}

@inproceedings{tang2026stealthrank,
  title = {{StealthRank}: {LLM} Ranking Manipulation via Stealthy Prompt Optimization},
  author = {Tang, Yiming and Fan, Yi and Yu, Chenxiao and Yang, Tiankai and Zhao, Yue and Hu, Xiyang},
  booktitle = {ICML 2026 Workshop on Trustworthy AI for Good},
  year = {2026},
  url = {https://openreview.net/forum?id=BTmGxDw32B}
}

@inproceedings{miroyan2026searcharena,
  title = {Search Arena: Analyzing Search-Augmented {LLMs}},
  author = {Miroyan, Mihran and Wu, Tsung-Han and King, Logan and Li, Tianle and Pan, Jiayi and Hu, Xinyan and Chiang, Wei-Lin and Angelopoulos, Anastasios and Darrell, Trevor and Norouzi, Narges and Gonzalez, Joseph E.},
  booktitle = {International Conference on Learning Representations},
  year = {2026},
  url = {https://proceedings.iclr.cc/paper_files/paper/2026/hash/4476dd7320e0eba63961990d73525064-Abstract-Conference.html}
}

@inproceedings{venkit2026deeptrace,
title={Deep{TRACE}: Auditing Deep Research {AI} Systems for Tracking Reliability Across Citations and Evidence},
author={Pranav Narayanan Venkit and Philippe Laban and Yilun Zhou and Kung-Hsiang Huang and Yixin Mao and Chien-Sheng Wu},
booktitle={The Fourteenth International Conference on Learning Representations},
year={2026},
url={https://openreview.net/forum?id=QkaeTea16Y}
}

@misc{deusser2026jev,
  title = {Evaluating and Benchmarking the System One Model Jev},
  author = {Deu{\ss}er, Tobias and Sparrenberg, Lorenz and Sifa, Rafet},
  year = {2026},
  eprint = {2609.37647},
  archivePrefix = {arXiv},
  primaryClass = {cs.CL},
  url = {https://arxiv.org/abs/2609.37647}
}

@misc{qwen2025qwen25technicalreport,
      title={Qwen2.5 Technical Report}, 
      author={{Qwen} and An Yang and Baosong Yang and Beichen Zhang and Binyuan Hui and Bo Zheng and Bowen Yu and Chengyuan Li and Dayiheng Liu and Fei Huang and Haoran Wei and Huan Lin and Jian Yang and Jianhong Tu and Jianwei Zhang and Jianxin Yang and Jiaxi Yang and Jingren Zhou and Junyang Lin and Kai Dang and Keming Lu and Keqin Bao and Kexin Yang and Le Yu and Mei Li and Mingfeng Xue and Pei Zhang and Qin Zhu and Rui Men and Runji Lin and Tianhao Li and Tianyi Tang and Tingyu Xia and Xingzhang Ren and Xuancheng Ren and Yang Fan and Yang Su and Yichang Zhang and Yu Wan and Yuqiong Liu and Zeyu Cui and Zhenru Zhang and Zihan Qiu},
      year={2025},
      eprint={2412.15115},
      archivePrefix={arXiv},
      primaryClass={cs.CL},
      url={https://arxiv.org/abs/2412.15115}, 
}

@inproceedings{
Chen2020TabFact:,
title={TabFact: A Large-scale Dataset for Table-based Fact Verification},
author={Wenhu Chen and Hongmin Wang and Jianshu Chen and Yunkai Zhang and Hong Wang and Shiyang Li and Xiyou Zhou and William Yang Wang},
booktitle={International Conference on Learning Representations},
year={2020},
url={https://openreview.net/forum?id=rkeJRhNYDH}
}

@inproceedings{
aly2021feverous,
title={{FEVEROUS}: Fact Extraction and {VER}ification Over Unstructured and Structured information},
author={Rami Aly and Zhijiang Guo and Michael Sejr Schlichtkrull and James Thorne and Andreas Vlachos and Christos Christodoulopoulos and Oana Cocarascu and Arpit Mittal},
booktitle={Thirty-fifth Conference on Neural Information Processing Systems Datasets and Benchmarks Track (Round 1)},
year={2021},
url={https://openreview.net/forum?id=h-flVCIlstW}
}

@inproceedings{lu-etal-2023-scitab,
    title = "{SCITAB}: A Challenging Benchmark for Compositional Reasoning and Claim Verification on Scientific Tables",
    author = "Lu, Xinyuan  and
      Pan, Liangming  and
      Liu, Qian  and
      Nakov, Preslav  and
      Kan, Min-Yen",
    editor = "Bouamor, Houda  and
      Pino, Juan  and
      Bali, Kalika",
    booktitle = "Proceedings of the 2023 Conference on Empirical Methods in Natural Language Processing",
    month = dec,
    year = "2023",
    address = "Singapore",
    publisher = "Association for Computational Linguistics",
    url = "https://aclanthology.org/2023.emnlp-main.483/",
    doi = "10.18653/v1/2023.emnlp-main.483",
    pages = "7787--7813"
}

\clearpage
\newpage
\appendix

\section{Evaluated Product Families}
\label{app:families}
\label{app:supplementary-tables}
Table~\ref{tab:families} lists the 18 product families of the 80-brand cohort and the number of brands in each.
\begin{table}[!htbp]
\centering
\caption{Product families in the 80-brand cohort (six products per category).}
\label{tab:families}
\small
\setlength{\tabcolsep}{4pt}
\begin{tabular}{@{}lr@{}}
\toprule
Family & $n$ \\
\midrule
Acoustic spaces & 6 \\
Floating and buoyant platforms & 3 \\
Air handling and filtration & 4 \\
Horticulture and propagation systems & 4 \\
Archival and conservation equipment & 6 \\
Kilns and thermal processing furnaces & 5 \\
Brewing and beverage implements & 4 \\
Lodging and retreat places & 5 \\
Cold and cold chain storage & 4 \\
Metrology and calibration instruments & 4 \\
Drying and dehydration equipment & 7 \\
Subterranean workspaces & 4 \\
Elevated lookout decks & 4 \\
Textile and material craft equipment & 6 \\
Embedded observation stations & 3 \\
Transport and handling gear & 2 \\
Fermentation and brewing vessels & 4 \\
Water treatment and distillation apparatus & 5 \\
\bottomrule
\end{tabular}
\par\smallskip\parbox{0.98\columnwidth}{\footnotesize }
\end{table}

\section{Additional Results}
\label{app:additional}

\begin{figure*}[!t]
  \centering
  \includegraphics[width=\textwidth]{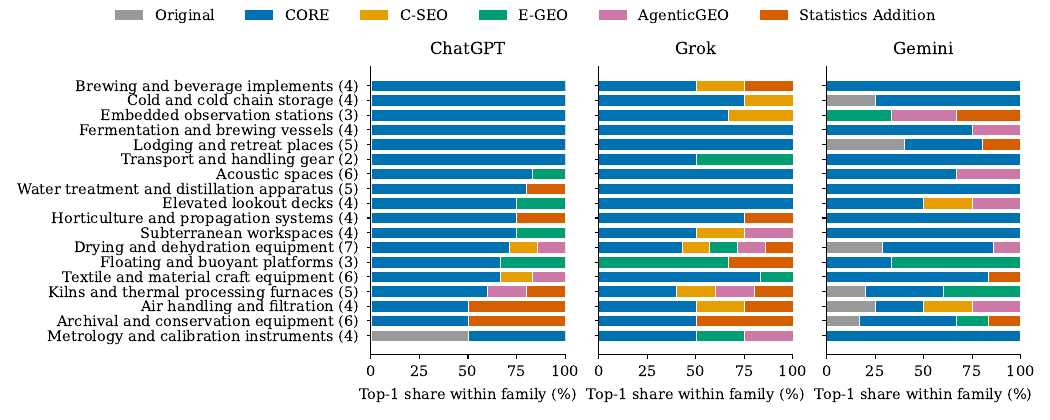}
  \caption{Share of brands in each product family whose top-ranked product comes from each strategy, for ChatGPT, Grok, and Gemini under frozen $\pm0.2\%$ prices (product URLs). Numbers in parentheses are brands per family.}
  \Description{Three panels of horizontal stacked bars, one per GE, showing for each of the 18 product families which strategy wins the Top-1 position and in what share of brands.}
  \label{fig:family-top1}
\end{figure*}

Figure~\ref{fig:agreement-heatmap} extends Table~\ref{tab:agreement} to all 28 evaluator pairs: the highest agreement among direct-text evaluators is between GPT-4o and DeepSeek ($\rho=0.62$), above any pair that involves a GE, and Claude correlates negatively with every other evaluator. Figure~\ref{fig:family-top1} breaks the Top-1 rates of Table~\ref{tab:ranks} down by product family for the three GEs under frozen $\pm0.2\%$ prices; families are ordered by CORE's share on ChatGPT. Table~\ref{tab:brand-winners} lists the top-ranked product of every brand for the three GEs and DeepSeek.

\begin{table*}[!t]
\centering
\caption{Top-ranked product of each brand under frozen $\pm0.2\%$ prices (GEs: product URLs; DeepSeek: direct text). Orig.\ = Original, Agentic = AgenticGEO, Stat.\,Add.\ = Statistics Addition.}
\label{tab:brand-winners}
\footnotesize
\setlength{\tabcolsep}{3pt}
\renewcommand{\arraystretch}{0.88}
\begin{tabular*}{\textwidth}{@{\extracolsep{\fill}}lllll@{\hspace{1.5em}}lllll@{}}
\toprule
Category & ChatGPT & Grok & Gemini & DeepSeek & Category & ChatGPT & Grok & Gemini & DeepSeek \\
\midrule
actinic shielded print cabinet & CORE & CORE & CORE & CORE & eutectic buffered dairy cellar & CORE & CORE & Orig. & CORE \\
active layer thaw bed planter & Stat.\,Add. & Stat.\,Add. & CORE & CORE & eutectic buffered film cold store & Stat.\,Add. & Stat.\,Add. & Stat.\,Add. & C-SEO \\
adit entry survey chamber & CORE & CORE & CORE & CORE & eutectoid stage hardening retort & Stat.\,Add. & Stat.\,Add. & CORE & CORE \\
advection fog demister & CORE & CORE & Agentic & CORE & evaporite mineral brining tun & CORE & CORE & Agentic & Orig. \\
aeolian sand culture tray & CORE & CORE & CORE & CORE & falling rate belt dryer & CORE & CORE & Orig. & CORE \\
aeolian sealed dripper & CORE & C-SEO & CORE & CORE & ferrite stage induction bath & CORE & CORE & E-GEO & CORE \\
ambient floor audition vault & CORE & CORE & Agentic & CORE & fetch limited pontoon stage & E-GEO & Stat.\,Add. & E-GEO & CORE \\
anchor ice launch cradle & CORE & E-GEO & CORE & CORE & firn core acoustic test cell & CORE & CORE & CORE & CORE \\
andosol mix rooting box & CORE & CORE & CORE & CORE & firn line bothy & CORE & CORE & Orig. & CORE \\
anorthosite block quench pit & CORE & CORE & CORE & CORE & firn probe calibration bench & Orig. & E-GEO & CORE & Agentic \\
anoxic parchment cabinet & Stat.\,Add. & Stat.\,Add. & CORE & Stat.\,Add. & firn sealed grain locker & CORE & CORE & CORE & C-SEO \\
anoxic sourdough crock & CORE & CORE & CORE & CORE & flaw lead sighting deck & CORE & E-GEO & CORE & CORE \\
ashfall shielded viewing deck & CORE & CORE & C-SEO & C-SEO & flint temper glost kiln & CORE & C-SEO & Orig. & CORE \\
aufeis jacketed cheese vault & CORE & C-SEO & CORE & E-GEO & foehn sheltered ridge lookout & E-GEO & CORE & Agentic & CORE \\
austenitic hold annealing oven & Agentic & Agentic & E-GEO & Stat.\,Add. & footwall drive fabrication bay & CORE & Agentic & CORE & CORE \\
barophilic pressure vat & CORE & CORE & CORE & CORE & fulvic bath scouring tank & CORE & CORE & CORE & CORE \\
basalt bed kettle & CORE & CORE & CORE & CORE & fumarole fed steeping urn & CORE & CORE & CORE & CORE \\
benthic range pressure calibrator & CORE & CORE & CORE & CORE & gas shielded fumarole station & CORE & C-SEO & Agentic & CORE \\
blockfield guest lodge & CORE & CORE & CORE & CORE & geothermal flash evaporator & CORE & CORE & CORE & E-GEO \\
bora rated cliff walkway & CORE & CORE & CORE & CORE & gibber plain collection hopper & Stat.\,Add. & CORE & Orig. & CORE \\
brine flat field retreat & CORE & CORE & Orig. & CORE & gilgai cycle test chamber & CORE & Agentic & CORE & CORE \\
brine mist fogponic rack & CORE & CORE & CORE & CORE & glass transition noodle dryer & CORE & E-GEO & CORE & CORE \\
caliche sealed pottery vault & CORE & CORE & CORE & CORE & gley bath steeping vat & CORE & CORE & CORE & CORE \\
chott air spice tray & C-SEO & C-SEO & Orig. & CORE & graded tephra collector & CORE & Agentic & CORE & CORE \\
clathrate cooled spirits hold & CORE & CORE & CORE & CORE & graded tephra slow filter & CORE & CORE & CORE & CORE \\
coincidence dip test chamber & CORE & CORE & CORE & CORE & graupel chilled infusion carafe & CORE & Stat.\,Add. & CORE & C-SEO \\
compacted loess speech room & E-GEO & CORE & CORE & CORE & gypcrete pack hardness filter & CORE & CORE & CORE & CORE \\
conidial load filter bank & Stat.\,Add. & C-SEO & C-SEO & CORE & haboob rated observation mast & CORE & CORE & CORE & CORE \\
convergence monitored office suite & CORE & CORE & CORE & Agentic & halocline dive lodge & CORE & CORE & CORE & CORE \\
cryoplanation terrace bunkhouse & CORE & CORE & Stat.\,Add. & CORE & halocline trimmed dive dolly & CORE & CORE & CORE & CORE \\
cryptoendolith shield stone case & CORE & CORE & Orig. & CORE & halophile culture brine tun & CORE & CORE & CORE & CORE \\
cryptogamic dust filter drum & CORE & Stat.\,Add. & CORE & CORE & halophile culture dye vat & CORE & CORE & Stat.\,Add. & CORE \\
desorption stage drying rack & CORE & Stat.\,Add. & CORE & CORE & hyaloclastite vent imaging post & CORE & CORE & Stat.\,Add. & Agentic \\
dilatant phase printing bench & CORE & E-GEO & CORE & CORE & hygroscopic trap film canister & Stat.\,Add. & Stat.\,Add. & E-GEO & CORE \\
directional freeze brine separator & Stat.\,Add. & CORE & CORE & CORE & hyporheic flow sand column & CORE & CORE & CORE & CORE \\
drift level machine gallery & E-GEO & C-SEO & CORE & CORE & impact damped rehearsal loft & CORE & CORE & CORE & CORE \\
dunite shell calender roller & Agentic & CORE & CORE & CORE & infrasound monitoring booth & CORE & CORE & Agentic & CORE \\
enthalpy wheel conditioning bay & Agentic & CORE & Agentic & CORE & infrasound pressure reference cell & Orig. & CORE & CORE & Stat.\,Add. \\
equilibrium moisture storage silo & CORE & CORE & CORE & CORE & isopiestic hold warping frame & C-SEO & CORE & CORE & CORE \\
estuarine mud viewing capsule & CORE & CORE & E-GEO & Orig. & jokulhlaup rated river stage & CORE & E-GEO & E-GEO & CORE \\
\bottomrule
\end{tabular*}
\end{table*}

\section{Identity Masking}
\label{app:masking}

In the masked condition, each brand's six strategy labels, including Original, are replaced by six neutral aliases (Luma, Tavi, Neri, Vela, Sori, and Miro), shuffled independently within each brand with a fixed seed; a product named \emph{Brand CORE}, for example, becomes \emph{Brand Vela}. The alias replaces the full product name and any standalone strategy label in titles, page metadata, links, descriptions, and reviews, while all other wording, the specifications, and the prices stay unchanged. Masked pages use opaque URL paths, navigation to unmasked pages is disabled, and the alias-to-strategy mapping is kept private.

\section{Ranking Prompts}
\label{app:prompts}

We adapt CORE's ranking prompt with small adjustments for LIVE-GEO's webpage evaluation setting. The product-URL template below reproduces the saved protocol with category and URL placeholders. The six URLs follow the frozen per-brand order. Opposite-order conditions change only their sequence; the masked condition substitutes the mapped URLs. Brand-URL requests replace the six-page introduction with ``Use the following brand page to discover its six candidate products:'' and supply one URL. The price requests additionally require a \texttt{displayed\_price} field in each reading-log entry. Neither prompt supplies previous ranks.

\begin{quote}\footnotesize\raggedright
You are a helpful assistant. The user query is:\\
best \{category\}\\
Use the following six candidate product pages:\\
\textless listing URL 1\textgreater\\
\dots\\
\textless listing URL 6\textgreater\\
Before ranking, open each individual product page and read its product description and all displayed customer comments. Do not judge from the brand-page summaries alone.\\
After reading, recommend the products by producing a ranked list from most to least relevant to the user query. Only use the information provided on these pages.\\
Return a JSON object containing:\\
1. ``ranking'': an array of the six exact product names, ordered from rank 1 (best) to rank 6 (worst).\\
2. ``reading\_log'': one entry per product, containing:\\
``product\_name''; ``product\_url''; ``description\_read'': ``full'', ``partial'', or ``none''; ``comments\_displayed'': the number of individual comments exposed by the page, or null if unknown; ``comments\_read'': the number actually read, or null if unknown; ``access\_limitation'': a brief description of any missing or inaccessible content, or null.\\
The aggregate review count is not the number of comments displayed. Do not treat it as the number read.\\
Report access and reading coverage honestly. Do not invent content, counts, or reading percentages.\\
Include a ``limitations'' field:\\
Use null if all six descriptions and all individually displayed comments were fully read. Otherwise, describe only the actual access gaps in one sentence of at most 40 words. A larger aggregate review count than the displayed comment count is not an access failure. Do not include ranking explanations or general disclaimers in this field.
\end{quote}

The direct-text evaluators receive one user message built from the same frozen records, with the six products as a JSON array:

\begin{quote}\footnotesize\raggedright
You are a helpful assistant. The user query is: best \{category\}\\
The candidate products are:\\
{[}one object per product with ``name'', ``price'', ``rating'', and ``description'' (the product description followed by the seven customer comments){]}\\
Recommend the products by producing a ranked list from most to least relevant to the user query.\\
Only use the provided information in the JSON input.\\
Return ONLY a JSON array of product names in ranked order, from rank 1 (best) to rank N (worst).\\
Example format: [``Product A'', ``Product B'', ``Product C'']
\end{quote}

\end{document}